\documentstyle[12pt]{article}
\begin{document}

\rightline{FTUV/94-34}

\begin{center}
{\LARGE{\bf A novel kind of neutrino oscillation experiment}}
\end{center}

\vspace{2cm}

\begin{center}
{\large{\bf J. Segura, J. Bernab\'{e}u, F.J. Botella and J. A. Pe\~{n}arrocha
}}

\vspace{1cm}

{\it Departament de F\'{\i}sica Te\`{o}rica \\ Universitat de Val\`{e}ncia \\
and \\ IFIC, Centre Mixt Univ. Val\`encia-CSIC \\ E-46100 Burjassot,
Spain}
\end{center}

\vspace{2cm}

\baselineskip 0.7cm

\begin{center}
{\bf Abstract}
\end{center}
{\noindent A novel method to look for neutrino oscillations is proposed
 based on the elastic scattering process $\bar{\nu}_{i} e^{-}
 \rightarrow \bar{\nu}_{i} e^{-}$, taking advantage of the dynamical zero
 present in the differential cross section for
 $\bar{\nu}_{e} e^{-}
 \rightarrow \bar{\nu}_{e} e^{-}$. An effective  tunable experiment
 between the "appearance" and "disappearance" limits is made possible.
 Prospects to exclude the allowed
 region for atmospheric neutrino oscillations are given.
}

\newpage

Three aspects are essential in a neutrino oscillation \cite{Pon}
 experiment: the source,
the evolution ( to allow oscillations ) and the detection. If we restrict
ourselves to flavour vacuum oscillations, essentially two types of experiments
have been proposed and performed : the so
 called appearance and disappearance \cite{Bil} limits. In an
appearance experiment a certain source produces a neutrino of a given flavour
and, after evolution, the experiment tries to detect neutrinos of another
flavour. Since flavour is defined through charged current interactions it is
customary to use a pure charged current interaction to detect the new flavour
. Charged current detection has a threshold for production so that it is
 impossible to use low energy electron neutrinos ( or antineutrinos ) for
 appearance experiments. In particular, it must be stressed that the copious
 reactor antineutrinos cannot be used for appearance experiments.
This is certainly a drawback to explore regions in the $\Delta m^{2}-
 sin^2 2\phi$ plane where both $\Delta m^2$ ( the difference of neutrino
 squared masses ) and $sin^2 2\phi$ ($\phi$ being the mixing angle) are
 small. Note that the use of low energy neutrinos implies a better
 sensitivity to
 low $\Delta m^2$ for a given distance from the source to the detector.
 In general, appearance experiments are more sensitive to small mixing
 angles than the disappearance ones.

   In disappearance experiments there is a controlled source which produces a
 given flavour and a detector which sees the same flavour via a charged
 current interaction with some target: a depletion in the
 flux after neutrino travelling would be
 a manifestation of oscillation. For small mixing, however,
 the dominant signal in the detector comes from neutrinos with the
 original flavour, so this translates into less sensitivity to small mixing
 angles in disappearance experiments.

   As a consequence we see that, for example,
 in order to explore in the laboratory the
 region of masses and mixing where potential neutrino
 oscillations from atmospheric neutrinos \cite{Hir}
 have been suggested it is necessary to consider detectors very far away
 from the source \cite{Par} with the corresponding reduction in the flux.

 For $\nu_{e}$ and $\bar{\nu}_{e}$ beams the previous arguments
are true provided the detection reaction is a pure charged current one.
 We are going to consider a  mixed charged
  and neutral current reaction
 like $\bar{\nu}_{e} e^{-} \rightarrow \bar{\nu}_{e} e^{-}$, making
 use of the fact that
 the corresponding
 cross section is different from the one for
 $\bar{\nu}_{\mu} e^{-} \rightarrow \bar{\nu}_{\mu} e^{-}$, which only has
 a neutral current contribution. In principle, it could be
 possible to perform a neutrino oscillation experiment ($\bar{\nu}_{e}
 \rightarrow \bar{\nu}_{\mu}$) just by measuring the cross section for the
 scattering of neutrinos on electrons at some distance from the neutrino
 source. If oscillations take place some component of the beam will change,
 for example, from $\bar{\nu}_{e}$ to $\bar{\nu}_{\mu}$, and the total
 number of recoil electrons detected will be different from the counting one
 would have if no oscillations occurred. However the fact that
 both total cross sections are similar
 ( at high energies the electron antineutrino total cross section is
 about 3 times larger that the corresponding one for muon antineutrino )
  seems to disfavour this possibility of studying oscillations.

   Nevertheless , recently it has been proved \cite{Seg} that the
 cross section for the scattering of electron antineutrinos on electrons
 has a dynamical zero for the kinematical configuration corresponding
 to maximum electron recoil energy $T$ for an incident antineutrino
 energy $E_{\nu}=m_{e}/(4sin^2 \theta_{W})\simeq m_{e}$, being $m_{e}$ the
 electron mass. This zero is not
 present in $\bar{\nu}_{\mu} e^{-} \rightarrow \bar{\nu}_{\mu} e^{-}$. So
 it is possible to imagine an experiment with an ideal monoenergetic beam
 with $E_{\nu}\simeq m_{e}$ measuring the cross sections with recoil
 electrons
 at $T=T_{max}=2E_{\nu}^2/(2E_{\nu}+m_{e})\simeq 2m_{e}/3$. If any signal is
found
 it would come from $\bar{\nu}_{\mu} e^{-} \rightarrow \bar{\nu}_{\mu} e^{-}$
 after the initial $\bar{\nu}_{e}$ has oscillated to $\bar{\nu}_{\mu}$.

   In the standard theory the differential cross section for the process
 $\bar{\nu}_{i} e^- \rightarrow \bar{\nu}_{i} e^-$ \cite{Vog} is given by

\begin{equation}
\frac{\displaystyle{d\sigma^{\bar{\nu}_{i}}}}{\displaystyle{dT}}=
\frac{\displaystyle{2 G^2 m_{e}}}{\displaystyle{\pi}}\left[
(g_{R}^i)^2+(g_{L}^i)^2\left(1-\frac{\displaystyle{T}}{\displaystyle{E_{\nu}}}
\right)^2
-g_{L}^i g_{R}^i \frac{\displaystyle{m_{e} T}}{\displaystyle{E_{\nu}^2}}\right]
\end{equation}

\noindent
where $G$ is the Fermi coupling constant, $T$ the recoil
 kinetic energy of the electron and $E_{\nu}$ the antineutrino incident energy.
 For $\nu_{i}$ one has to make the change $g_{L}^i\leftrightarrow g_{R}^i$
 in Eq. (1).
 In terms of the weak mixing angle $\theta_{W}$, the chiral couplings
 $g_{L}^i$ and $g_{R}^i$ can be written for each neutrino flavour as

\begin{equation}
\begin{array}{cc}
g_{L}^e=\frac{1}{2} + sin^2\theta_{W} , & g_{R}^e=sin^2\theta_{W}\\
\\
g_{L}^{\mu ,\tau}=-\frac{1}{2} + sin^2\theta_{W} ,
& g_{R}^{\mu ,\tau}=sin^2\theta_{W}
\end{array}
\end{equation}

{}From Eq. (1) it is evident that if $g_{L}^i g_{R}^i>0$ there is a chance
 for the cross section to cancel in the physical region. From Eq. (2)
 we see that this zero is only possible in the $\bar{\nu}_{e} e^-
 \rightarrow \bar{\nu}_{e} e^-$ channel and, in fact, it takes place for the
 kinematical configuration $E_{\nu}=m_{e}/(4 sin^2\theta_{W})$ and
 maximal $T$. Neither
 $d\sigma^{\bar{\nu}_{\mu}}/dT$ nor $d\sigma^{\bar{\nu}_{\tau}}/dT$
 present a dynamical zero since $g_{L}^{\mu ,\tau}g_{R}^{\mu ,\tau}<0$.
 We will take advantage of this fact to propose a novel kind of neutrino
 oscillation experiment for reactor antineutrinos.
 It is important to stress several additional facts which explain why it is
 worthwhile to study more carefully this sort of "appearance" experiment using
 a neutral current reaction for neutrino "detection":

\begin{description}
\item{i)} The dynamical zero is only present for $\bar{\nu}_{e}$, not for
$\nu_{e}$ or $\nu_{\mu}$ ($\bar{\nu}_{\mu}$), $\nu_{\tau}$
($\bar{\nu}_{\tau}$).

\item{ii)} The flavour $\bar{\nu}_{e}$ is precisely the one which is produced
copiously in nuclear reactors.

\item{iii)} The neutrino energy  at which the zero appears
is on the peak of the
 antineutrino reactor spectrum \cite{Vog,Kla}.

\item{iv)}  The dynamical zero is located at the maximum electron recoil energy
$T\simeq 2m_{e}/3$. This value is in the range of the proposed
 experiments \cite{Bro,Gia,Bon} to detect recoil electrons.

\item{v)} In spite of the fact that the
 antineutrino reactor spectrum is continuous,
 there are several proposals with detectors that will be able to select the
 incident neutrino energies by measuring both
 the electron recoil energy and its
 recoil angle

\item{vi)} Last but not least, the use of 0.5 MeV reactor antineutrinos
 in this sort of appearance experiment would imply a good
 sensitivity to rather
 low $\Delta m^2$ values. In particular, as we will see, this kind of
 experiment would be sensitive to the region in the $\Delta m^2 -
 sin^2 2\phi$ plane where potential oscillations have been suggested
 from atmospheric neutrinos.

\end{description}

  The purpose of this paper consists in studying the potentialities of
 using the detection reaction $\bar{\nu}_{i} e^{-} \rightarrow
 \bar{\nu}_{i} e^{-}$ in a neutrino oscillation experiment, taking advantage
 of the dynamical zero present in the cross section for
 $\bar{\nu}_{e} e^{-} \rightarrow
 \bar{\nu}_{e} e^{-}$.

Suppose we have a source of electron-antineutrinos $\bar{\nu}_{e}(0)$ ( a
 nuclear reactor for example ) and we measure the differential cross
 section for the process $\bar{\nu}_{e}(x) e^- \rightarrow \bar{\nu}_{e}(x)
 e^-$ at a distance $x$ from the source. If vacuum oscillations take
 place we will have

\begin{equation}
\frac{\displaystyle{d\sigma^{\bar{\nu}}(E_{\nu},T,x)}}{\displaystyle{dT}}=
P_{\bar{\nu}_{e}\rightarrow \bar{\nu}_{e}}(x)
\frac{\displaystyle{d\sigma^{\bar{\nu}_{e}}(E_{\nu},T)}}{\displaystyle{dT}}
+\sum_{i=\mu,\tau}P_{\bar{\nu}_{e}\rightarrow \bar{\nu}_{i}}(x)
\frac{\displaystyle{d\sigma^{\bar{\nu}_{i}}(E_{\nu},T)}}{\displaystyle{dT}}
\end{equation}

\noindent
where $P_{\bar{\nu}_{e}\rightarrow \bar{\nu}_{i}}(x)$ is the probability of
 getting a $\bar{\nu}_{i}$ at a distance $x$ form the source. Taking advantage
 of the conservation of probability ( we disregard oscillation to
 sterile neutrinos ) and the identity
$d\sigma^{\bar{\nu}_{\mu}}
/dT=d\sigma^{\bar{\nu}_{\tau}}/dT$, Eq. (3) can be written as

\begin{equation}
\frac{\displaystyle{d\sigma^{\bar{\nu}}(E_{\nu},T,x)}}{\displaystyle{dT}}=
\frac{\displaystyle{d\sigma^{\bar{\nu}_{e}}(E_{\nu},T)}}{\displaystyle{dT}}
+
\left(\frac{\displaystyle{d\sigma^{\bar{\nu}_{\mu}}
(E_{\nu},T)}}{\displaystyle{dT}}
-\frac{\displaystyle{d\sigma^{\bar{\nu}_{e}}(E_{\nu},T)}}{\displaystyle{dT}}
\right)\sum_{i=\mu,\tau}P_{\bar{\nu}_{e}\rightarrow \bar{\nu}_{i}}(x)
\end{equation}

In the particular case of considering only two flavour oscillation we have:

\begin{equation}
\sum_{i=\mu,\tau}P_{\bar{\nu}_{e}\rightarrow \bar{\nu}_{i}}(x)
\rightarrow P_{\bar{\nu}_{e}\rightarrow \bar{\nu}_{\mu}}(x)=
sin^2 2\phi sin^2 \left(\frac{\displaystyle{\Delta m^2 x}}{
\displaystyle{4 E_{\nu}}}\right)
\end{equation}

\noindent
where $\phi$ is the vacuum mixing angle  and $\Delta m^2$ is the difference
 of the square of masses of the mass eigenstates $\nu_{1}$
 and $\nu_{2}$.

 From Eq. (4) it is quite evident that by measuring $d\sigma^{\bar{\nu}}
/dT$ at the kinematical configuration where $d\sigma^{\bar{\nu}_{e}}
/dT$ vanishes, the signal will be proportional to the oscillation probability
 times the $\bar{\nu}_{\mu} e^- \rightarrow \bar{\nu}_{\mu} e^-$ cross
 section thus simulating  an "appearance" experiment. Moving to different
 configurations one has the possibility to tune the relative contribution
 of the two terms of Eq.(4) and thus the sensitivity to $sin^2 2\phi$
 and $\Delta m^2$.
 It must be stressed that Eq. (4) is a valid description for the
 neutrino-electron cross section
 for any energy $E_{\nu}$ and $T$ and so in principle
 it could be used with any source of electron antineutrinos. Advantage
 must be taken from this dependence on $E_{\nu}$ and $T$ in order to
 select the appropriate kinematical region in each
 type of experiment. On top of the dynamical zero our observable
  is a pure "appearance" experiment ,outside the zero it is a mixture
  of "appearance" and "disappearance"; the observable could simulate
  a pure disappearance experiment if there were regions where
  $\frac{d\sigma^{\bar{\nu}_{e}}}{dT}>>\frac{d\sigma^{\bar{\nu}_{\mu}}}{dT}$
  but this is not the case.

  With high energy neutrinos $(E_{\nu}>>m_{e})$, the $\bar{\nu_{e}}$
 cross section is larger that the $\bar{\nu_{\mu}}$ and, in addition,
 semileptonic charged current reactions
 ( when possible) are much more copious. So, unless a great precision
 could be reached, the use of these detection reactions with high energy
 neutrinos does not look the most appealing.

 Therefore we shall concentrate in studying
 the potentialities of the observable
 described in Eq. (4) near the dynamical zero for electron antineutrinos
 . This in turn means we will
 restrict ourselves to electron antineutrino beams coming from nuclear
 reactors.

 In order to select the appropriate neutrino energy $E_{\nu}$ from a
 continuous spectrum, it is necessary to measure the electron recoil angle
 $\theta$ and the electron recoil energy $T$.
 To a different extent, this can be done in the proposed experiments
 such as MUNU \cite{Bro}, BOREXINO \cite{Gia} and HELLAZ \cite{Bon}.
 The kinematical relation to be used is

\begin{equation}
cos\theta =\left. \frac{\displaystyle{T}}
{\displaystyle{\sqrt{T^2+2m_{e}T}}}\right(1+\left.
\frac{\displaystyle{m_{e}}}{\displaystyle
{E_{\nu}}}\right)
\end{equation}

With the aim of analyzing the more convenient phase space regions
for the measurement of Eq. (4) we present in Fig. 1 the curves for
 constant values of $d\equiv log\left[\frac{d\sigma^{\bar{\nu}_{\mu}}}{dT}
 /\frac{d\sigma^{\bar{\nu}_{e}}}{dT}\right]$ (solid lines)
 in the plane $(\theta,T)$.
Of course the curve $d=\infty$ collapses to the point $T\simeq 2m_{e}/3$
 and $\theta =0$, equivalent to $E_{\nu}\simeq m_{e}$
 and  $T=T_{max}\simeq 2m_{e}/3$.

{}From Fig. 1 it is evident that the width of the peak in the ratio
 $\frac{d\sigma^{\bar{\nu}_{\mu}}}{dT}/\frac{d\sigma^{\bar{\nu}_{e}}}{dT}$
 owing to the dynamical zero is relatively large in both directions $\theta$
 and $T$. For example in the window $0.2<T<0.6$ $MeV$ and $\theta <0.2$ $rad$
 the muon antineutrino cross section is at least 4 times bigger than the
 corresponding electron-antineutrino one. This kind of window enters in the
 capabilities of the previously mentioned experiments.

 In Fig. 1, we have also plotted (dashed lines) the curves of constant
 $E_{\nu}$. As we will show later these lines will be useful to cancel
 uncertainties coming from a poor knowledge
 of the antineutrino spectrum \cite{Vog,Kla}.

 For the moment, let us suppose that the
 antineutrino spectrum from the reactor is known.
 In Fig. 2 we present a plausible neutrino spectrum
  to be used in the rest of the paper .

  By inspection of Fig. 1 and Eq. 4, we see that our observable
  will be more sensitive to $\Delta m^2$ and $\phi$ in regions where
 $d$ is bigger, so the strategy we propose is to select events inside a region
 where $d$ is bigger than a certain number.

   As an illustration, if we calculate the number of
 events in terms of $\Delta m^2$ and $\phi$ in the region $d>log(5)
\simeq 0.7$, take the ratio to the number of events in the absence
 of oscillations and impose this ratio to be $1\pm 0.5$, the would-be
 exclusion plot
 we get is represented in Fig. 3.
 We have considered that the detector is placed 20 meters away from
 the reactor. Inside the excluded region we have inserted the by now allowed
 region of oscillations coming from atmospheric neutrino experiments. Taking
 into account the original MUNU proposal, the numbers we have considered
 correspond roughly to detect a few ($\sim 10$) events per year if no
 oscillations take place.

  If instead of taking the window $d>log(5)$ we consider bigger windows,
 $d>log(4)$ for example, there are two effects which operate in opposite
 sense. Let us
 suppose that the precision of the ratio $\#$ (events with oscillations)/
 $\#$ (events without oscillations) goes as $1/\sqrt{N}$, where $N$ is the
 number of detected events. Then if we integrate over $d>log(4)$ we will have
 more events than with $d>log(5)$ and so higher precision. However,
 we are
 including regions where $d$ is smaller in such a way that the second piece
 in the right hand side of Eq. (4) is less important and so the observable
 defined in this equation will be less sensitive to $\Delta m^2$ and $\phi$.
 If only these facts are taken into account we have checked that in order to
 exclude the atmospheric neutrino region one gets similar results if the
 window chosen is $d>log(1), \,\, log(2),\,\, ... log(5)$. We do not consider
 higher $d$ to avoid having too few events.

  These results clearly show the feasibility of this kind of
experiment with the proposed detectors previously mentioned and establish
 the interest of the proposal in order to study the region in the
 $\Delta m^2 -\phi$ plane where potential neutrino oscillations have
 been reported from atmospheric neutrino data.
 These conclusions have been obtained under the assumption of a
 known neutrino flux. Note that the total number of events in a given
 kinematical region depends on both the weak cross section at the distance
 $x$ and the antineutrino spectrum. So, in order to get a
 precision measurement of the cross section it is necessary either to
 measure the antineutrino spectrum
 in the same experiment or to reduce to a minimum the uncertainties coming
 from the poor knowledge of the neutrino spectrum, specially around
 $E_{\nu}\simeq 0.55 MeV$.

   The way to measure the neutrino flux in the same experiment looks
 theoretically
 very simple. One just has to measure the number of events in a kinematical
 region where the dependence on $\Delta m^2$ and $\phi$ disappears. Looking
 at Eq. (4) it is evident that this kinematical region is precisely the region
 where $d\sigma^{\bar{\nu}_{\mu}}/dT=d\sigma^{\bar{\nu}_{e}}/dT$ which
 corresponds to the curve $d=log(1)=0$ in Fig. 1. So we propose to measure
 the number of events in the region $d=0$ where the cross section is given
  by

\begin{equation}
\left.\frac{\displaystyle{d\sigma^{\bar{\nu}}}}{\displaystyle{dT}}(E_{\nu},T,x)
\right|_{d=0}=
\left.\frac{\displaystyle{d\sigma^{\bar{\nu}_{e}}}}
{\displaystyle{dT}}(E_{\nu},T)
\right|_{d=0}
\end{equation}

  So by dividing the number of events in the region $d=0$ by the well known
 cross section in Eq. (7) one gets directly the neutrino flux. It is
 interesting to note that the cross section in Eq. (7) is also independent of
 $x$ and so it could be very useful in the calibration of the detectors
 in an experiment with two detectors placed at different values of $x$.

  Depending on the actual setup of an experiment, it may be that
 only a few
 events around the $d=0$ zone could be detected; then, it would be necessary
 to use all the available information and not just the events around $d=0$.
 So let us construct the following observable

\begin{equation}
R_{c}=\frac{\displaystyle{N(c)}}{\displaystyle{N(\Delta c)}}
\end{equation}

\noindent
where $N(c)$ is the number of detected events in the region $d>log(c)$
 and $N(\Delta c)$ the total number of events inside the  region
 $0<\theta <0.5$ radians and $E_{\nu}^{min}(c)<
 E_{\nu}<E_{\nu}^{max}(c)$ where $E_{\nu}^{min}(c)$ and $E_{\nu}^{max}(c)$ are
 the minimum and maximum values of $E_{\nu}$ for the boundary region
 $d\geq log(c)$. By choosing the $E_{\nu}$ boundaries of $\Delta c$ in this
way,
 the same part of the spectrum enters in the numerator
 and in the denominator of $R_{c}$ in such a way that the uncertainties
 tend to cancel. The $\theta$ boundary in $\Delta c$ could be chosen
 ,depending on the final setup of the experiment, in a different way.
 We have checked that when we change the spectrum of Fig. 2 by a $50\%$
 in the region $E_{\nu} \leq 1.5\, MeV$ the error introduced in
 $R_{c}$ is, for $c=5$, of order $10\%$ and so
 much smaller than the $50\%$ precision needed
 to draw the exclusion plot in Fig. 3.
 For $c=5$ the window in $T$ corresponds to the reasonable
 range $0.1<T<1.25 MeV$.
   So we can conclude that without including any  systematic errors such
 as geometrical acceptances and so on, the uncertainties coming from the
 neutrino spectrum can be put under control either by measuring the spectrum
 or by the use of ratios of the kind presented in \mbox{Eq. (8)}.

In conclusion, we have presented a novel
 kind of neutrino oscillation experiment
 based on the neutral current reaction $\bar{\nu}_{i} e^- \rightarrow
 \bar{\nu}_{i} e^-$, where advantage from the dynamical zero present in the
 cross section for $\bar{\nu}_{e} e^- \rightarrow \bar{\nu}_{e} e^-$ has
 been  taken.
 This kind of experiment is in general a mixture of "appearance" and
 "disappearance" experiments. For antineutrinos coming from a nuclear power
 plant and using detectors capable of measuring both the recoil
 energy of the electron and its recoil angle (MUNU, BOREXINO, HELLAZ )
 this kind of experiment could in principle be sensitive to the by now
 allowed region of neutrino oscillation coming from atmospheric neutrino
 data.

{\bf ACKNOWLEDGEMENTS}

\hspace{0.5cm}This paper has been supported by CICYT under Grant AEN 93-0234
 . We are
 indebted to G. Bonvicini, C. Broggini, J. Busto,
J.F. Cavaignac, M. Giammarchi and
 D.H. Koang
for discussions about the topic of this paper.

\newpage
\begin{center}
{\large {\bf Figure Captions.}}
\end{center}

\begin{itemize}
     \item{{\bf Fig. 1)} Curves for constant values of $d\equiv log[\frac{
d\sigma^{\bar{\nu}_{\mu}}}{dT}/\frac{
d\sigma^{\bar{\nu}_{e}}}{dT}]$ (solid lines), from d=0 to d=2, and for
 constant values of $E_{\nu}$ in MeV (dashed lines) in the plane $(T,\theta)$.}
     \item{{\bf Fig. 2)} A plausible $\bar{\nu}_{e}$ spectrum from a nuclear
 reactor.}
      \item{{\bf Fig. 3)}
  Would-be exclusion plot
 obtained by imposing that the ratio
 $\int \frac{d\sigma^{\bar{\nu}}}{dT}
/\int\frac{d\sigma^{\bar{\nu}_{e}}}{dT}$ (ratio oscillation/non-oscillation)
is less than $1.5$,
integrating the cross sections over
 a typical reactor spectrum in the kinematical
 region where $\frac{d\sigma_{\bar{\nu}_{\mu}}}{dT}/
\frac{d\sigma_{\bar{\nu}_{e}}}{dT}\geq 5$ and considering the detector is
 $20$ meters away from the reactor. The shaded zone corresponds to the allowed
 region for atmospheric $\nu_{e}\leftrightarrow\nu_{\mu}$ oscillations.}

\end{itemize}

\end{document}